\documentstyle[aps,prl,twocolumn,epsfig] {revtex}
\begin{document}

\title{Time evolution of thermodynamic entropy for conservative \\and 
dissipative chaotic maps}
 
\author{M. Baranger$^a$, V. Latora$^b$ and A. Rapisarda$^b$}

\address{$^a$Center for Theoretical Physics, Laboratory for Nuclear 
Sciences and  Department of Physics, 
\\Massachusetts Institute of Technology, Cambridge, Massachusetts 
02139, USA\\}
 
\address{$^b$ Dipartimento di Fisica e Astronomia Universit\'a di 
Catania and INFN sezione di Catania,\\
Corso Italia 57, I-95129 Catania, Italy}

\date{\today}
\maketitle

\begin{abstract}
We consider several  low--dimensional chaotic maps 
started in far-from-equilibrium initial conditions and we study 
the process of relaxation to equilibrium.
In the case of conservative maps the Boltzmann-Gibbs entropy S(t) 
increases linearly in time with a slope equal to the Kolmogorov-Sinai 
entropy rate.  The same result is obtained also for a simple case of 
dissipative system, the logistic map, when considered in the chaotic 
regime.  
A very interesting results is found at the chaos threshold. In this case, 
the usual Boltzmann-Gibbs is not 
appropriate and  in order to have a linear increase, as for the chaotic case, 
we need to use the  generalized q-dependent Tsallis entropy $S_q(t)$
with a particular value of a q different from 1 (when q=1 
the generalized entropy reduces to the Boltzmann-Gibbs). The entropic index 
q appears to be  characteristic of the dynamical system.  
\end{abstract}

%\begin{keyword}
%Hamiltonian dynamics, deterministic chaos, relaxation to equilibrium, 
%Kolmogorov entropy\\
%{\em PACS numbers:}~05.45.Pq, 05.20.-y, 05.60.cd, 05.70.Fh
%\end{keyword}

\section{Introduction}
In this paper we study the connection between two quantities, 
both called entropies and used in two different fields: 
the entropy of a thermodynamic system $S$  
(the entropy of Boltzmann and Clausius, the entropy of the 
second law of thermodynamics) \cite{balian}, 
and $\kappa$, the Kolmogorov--Sinai entropy, defined by the
mathematicians as a measure of chaos and describing the dynamical 
instabilities of trajectories in the phase space \cite{kolmo,beck}. 
As a main difference the thermodynamic entropy is a function of time, 
depending not only on the particular dynamical system, but also 
on the choice of an initial probability distribution for the state of
that system, while $\kappa$ is a single number, a property
solely of the chaotic dynamical system considered. 
Moreover, the Kolmogorov--Sinai entropy it is not really an entropy 
but an entropy--rate, i.e. an entropy per unit time.

The connection between these two quantities has not been addressed 
extensively in the literature. Only few and often very vague 
statements are present in the textbooks \cite{zaslav}. 
Though some similar ideas have appeared previously in the 
literature \cite{dellago,gu}, a clarification of this connection 
has been proposed for strongly chaotic hamiltonian systems by two of the 
authors in ref. \cite{baranger1}. 

Here we consider conservative and non--conservative chaotic systems. 
We focus in particular on two two--dimensional conservative maps, 
the cat and the standard map, and on the logistic map, the simplest 
one--dimensional dissipative system.  
We start the system in far--from--equilibrium initial conditions, 
and we follow numerically the process of relaxation to equilibrium.
When the dynamics is chaotic the variation with 
time of the physical entropy goes through 
three successive, roughly separated stages. 
In the first one, $S(t)$ is dependent on the details 
of the dynamical system and of the initial distribution, and 
no generic statement can be made. 
In the second stage, $S(t)$ is a linear increasing function of time. 
In the third stage, $S(t)$ tends asymptotically towards the 
constant value which characterizes equilibrium, for which the distribution 
is uniform in the available part of phase space.  
The actual connection requires $S(t)$ to be averaged over 
many histories, so as to give equal weights to initial distributions 
from all regions of phase space. 
When such average is performed a perfect equality $dS/dt= \kappa$ 
is obtained in the intermediate stage. 

In this paper we also address a very special situation, i.e. the case 
of the logistic map at the edge of chaos.  It has been shown that 
in order to be treated such a special border of chaos situation 
requires the nonextensive definition of  entropy $S_q(t)$
 \cite{barangertsallis}.

\section{Conservative Chaotic Systems}

As a first example we consider the 
``generalized cat map'', defined by the 
following iterative rule inside a unit square: 
\begin{eqnarray}
\label{cat1}
P &=& p  + aq     ~~~~~~~~\pmod 1~\nonumber \\
Q &=& p  +(1+a)q   ~\pmod 1~
\end{eqnarray}
where $a$ is a positive control parameter. 
In fig.~1 we consider an initial distribution 
very strongly localized in a tiny region of the phase space, 
and we let it evolve in time according 
to the eqs.~(\ref{cat1}) with $a=1$. 
The distribution stretches in one direction 
and contracts in another. The volume of the distribution in 
phase space is conserved because of the Liouville's theorem, 
while its shape keeps changing in time until the system 
reaches a stationary state with an uniform filling of  
the phase space.  

%%%%%%%%%%%%%%%%%%%%%%%%%%%%%%%%%%%%%%%%%%%%%%%%%%%%%%%%%%%%%%%%%%%%%%%%%%%
%%%%%%%%%  FIG. 1
\begin{figure}
\begin{center}
\epsfig{figure=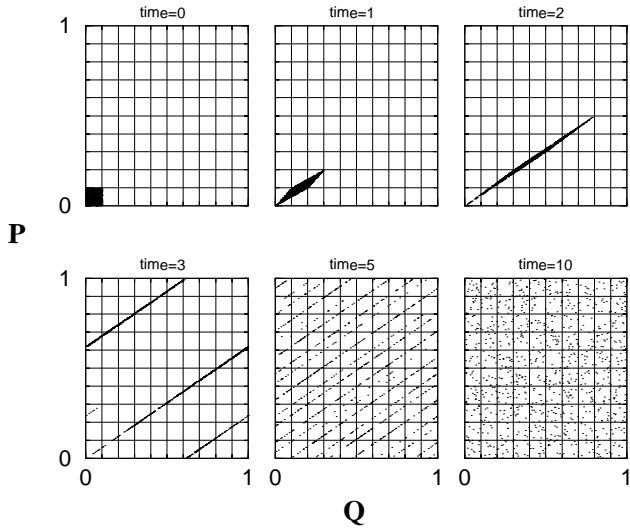,width=7truecm,angle=270}
\end{center}
\caption{ Generalized Cat Map with $a=1$. Time evolution of an initial 
distribution of points; we report the following time steps 0,1,2,3,5,10.}
\end{figure}
%%%%%%%%%%%%%%%%%%%%%%%%%%%%%%%%%%%%%%%%%%%%%%%%%%%%%%%%%%
%%%%%%%%%%%%%%%%%

Any fine--grained quantity (for example an entropy defined as an 
integral over the phase space) does not vary with time at all.   
However the shape of the volume  becomes increasingly 
complicated due to the chaotic dynamics. 
In order to have an entropy which is increasing in time 
(in agreement with the second principle of  thermodynamics)  
we simply need to perform a coarse--graining, i.e.
a slight smearing, or smoothing, 
of the probability distribution in phase space before calculating 
$S$ \cite{coarse}. 
There are many ways to introduce a coarse--graining. 
In this paper, we assume that phase space is divided into 
a grid of a large number $M$ of cells $c_{i}$ with volumes 
$v_{i}$, such that ${\sum}_{i} v_{i}=V$, 
the total volume of available 
phase space. 
In fig.~1 we report a grid of  $M=10$x$10$ cells, though for the actual 
calculations of this paper 
we will use a much more refined coarse--graining. 
Our definition for the out--of--equilibrium entropy is then the 
coarse--grained Boltzmann-Gibbs entropy: 
\begin{equation}
\label{scoarse}
S(t)  =  -    \sum_{i} ~  p_{i}(t) ~ 
             \log  { p_{i}(t)} ~~, 
\end{equation}
where $p_{i}(t)$ is the probability that the state 
of the system  falls inside 
cell $c_{i}$ of phase space  at time $t$.
Such a coarse--grained entropy is a quantity which is increasing 
in time, and the chaotic dynamics is the main reason of this increase. 
In the following of this paper we will show the 
fundamental importance of chaos in the relaxation to equilibrium: 
the entropy increases with a rate exactly equal to 
the Kolmogorov--Sinai entropy, the measure of chaos.  
N.S.Krylov, already in the 1940's \cite{krylov}, was
the first one who understood clearly that the {\it mixing} 
property of
chaos was essential for the foundations of statistical mechanics.
In fig.~2 we show $S(t)$ for four values of $a$ (see caption).
Now the coarse--graining grid is obtained by 
dividing each axis into 400 equal segments. 
The initial distribution consists of $10^6$ points placed at 
random inside a square whose size is that
of a coarse--graining cell, and the center of that square is 
picked at random anywhere on the map.
In this way the initial distribution is 
very strongly localized in phase space i.e. 
all the cells but a few contain initial zero probability. 

Each of the four curves is an 
average over 100 runs, i.e. 100 histories with different initial 
distributions chosen at random, as mentioned. Each curve shows clearly 
the stage--2 linear behavior, the slope $dS/dt$ being 
accurately given by the KS entropy--rate $\kappa$.
To calculate $\kappa$  here, we have used the fact that it is 
equal to the sum of the positive Lyapunov exponents \cite{pesin}. 
In the case of cat map we have only one  positive 
analytically calculable Lyapunov exponent $\lambda$, 
given by the equation:   
\begin{equation}
\label{cat2}
\kappa = \lambda= \log \frac{1}{2}({2 +a +\sqrt{a^2+4a}})~~. 
\end{equation}
The values of $\kappa$ are reported in the figure caption. 
%%%%%%%%%%%%%%%%%%%%%%%%%%%%%%%%%%%%%%%%%%%
%%%%%%%%%%%%%%%%%%%%%%%%%%%%%%%%
%%%%%%%%%  FIG. 2
\begin{figure}
\begin{center}
\epsfig{figure=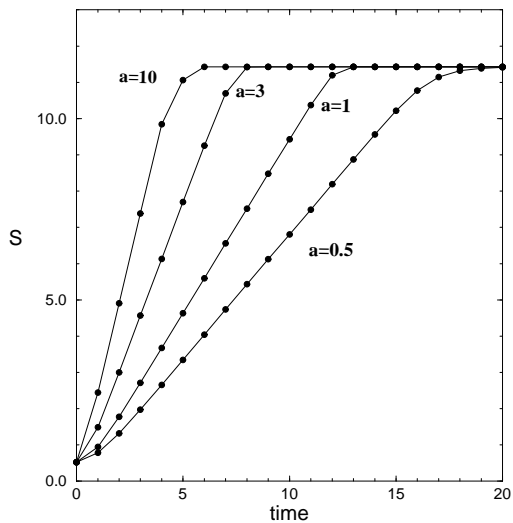,width=7truecm,angle=270}
\end{center}
\caption{ We show the thermodynamical entropy $S$ vs time for the 
Generalized Cat Map, with ~$a=10,3,1,0.5$~. We considered 
~$N=10^6$,
~grid $=400\times400$,
~$V_i=V_{\rm cell}$, and an 
average of 100 histories. The slope of the linear rise coincides with the
Kolmogorov-Sinai entropy which is 
~$\kappa=2.48,1.57,0.96,0.69$ respectively from right to left.
See text for more details.
}
\end{figure}
%%%%%%%%%%%%%%%%%%%%%%%%%%%%%%%%%%%%%%%%%%%%%%%%%%%%%%%%%%%%%%%%%%%%%%%%%%%
The equality $dS/dt=\kappa$ is valid in the intermediate 
stage for each of the four $a$ values considered in figure. 
Moreover it can be shown that this result does not depend 
on the size of the initial distribution and on the 
size of the coarse--graining cells \cite{baranger1}.  
The type of coarse--graining we adopted allows an alternative 
version of the significance of $\kappa$ for the evolution of a 
physical system.
Starting from an initial distribution localized in phase space 
(like in fig.1), during the generic second stage mentioned earlier, 
the total number of occupied cells, i.e. cells with
non-vanishing $p_{i}$, varies proportionally 
to $e^{\kappa t}$. 
Our simulations verify this fact well \cite{baranger1}. 

%%%%%%%%%%%%%%%%%%%%%%%%%%%%%%%%%%%%%%%%%%%%%%%%%%%%%%%%%%%%%%%%%%%%%%%%%%%
%%%%%%%%%  FIG. 3
\begin{figure}
\begin{center}
\epsfig{figure=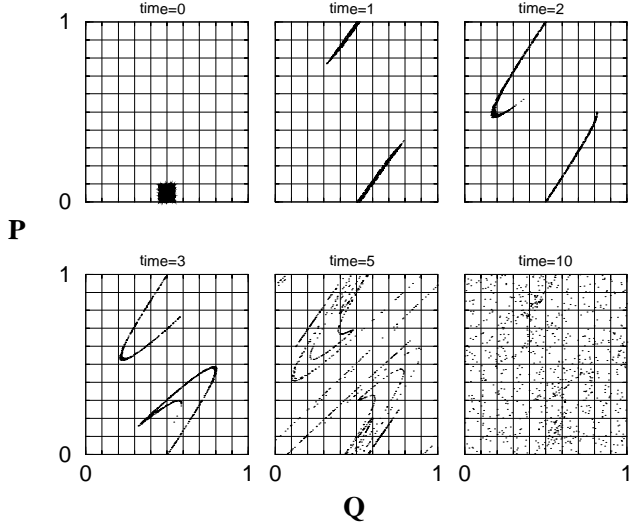,width=7truecm,angle=270}
\end{center}
\caption{ Standard Map with $a=5$. Time evolution of an initial 
distribution of points; we report time=0,1,2,3,5,10}
\end{figure}
%%%%%%%%%%%%%%%%%%%%%%%%%%%%%%%%%%%%%%%%%%%%%%%%%%%%%%%%%%%%%%%%%%%%%%%%%%%

The second system we have studied is the standard map \cite{stand}, again a 
two-dimensional conservative map in the unit square, but this time 
nonlinear:
\begin{eqnarray}
\label{sta}
P&=& p + \frac{a}{2\pi}~\sin(2\pi q) ~~~\pmod 1~ \nonumber, \\
Q&=& q + P                           ~~~~~~~~~~~~~~~~\pmod 1~. 
\end{eqnarray}
The map is only partially chaotic, but the percentage of chaos increases 
with the control parameter $a$, and we consider 
large values of $a$, namely 
20, 10, and 5. For $a=5$ there are still two sizeable regular islands, 
associated with a period 2 stable trajectory while for $a=10$ and $a=20$ 
the phase space is completely chaotic. 
In fig.~3 we report for the case $a=5$ the time evolution 
of an initial small distribution located in the chaotic region 
of the phase space. As a main difference with respect to fig.~1, is that for a 
nonlinear system the stretching and contraction rates and directions vary 
appreciably in the phase space from place to place.
The distribution bends on itself many times, nevertheless 
at the final time the system tends to occupy the whole chaotic part 
of the phase space in an uniform way.     
Fig.~4 presents three single histories (full lines)
~for $a=5$, as well as an average curve (circles) as in fig~.2.
The coarse--graining grid, the choice of initial distribution, 
and the averaging are the same used for the cat map, 
but it was necessary to include 1000 histories in order to obtain a 
good averaging. In fact, for a very nonlinear system as the standard 
map, the single curves can vary wildly and, differently from the case 
of the cat map, thus the averaging procedure is essential.  
Fig.~5 shows the final average curves $S(t)$ for three values $a$: 
the one at the bottom  corresponds to the smallest one. 
%%%%%%%%%%%%%%%%%%%%%%%%%%%%%%%%%%%%%%%%%%%%%%%%%%%%%%
%%%%%%%%%%%%%%%%%%%%%
%%%%%%%%%  FIG. 4
\begin{figure}
\begin{center}
\epsfig{figure=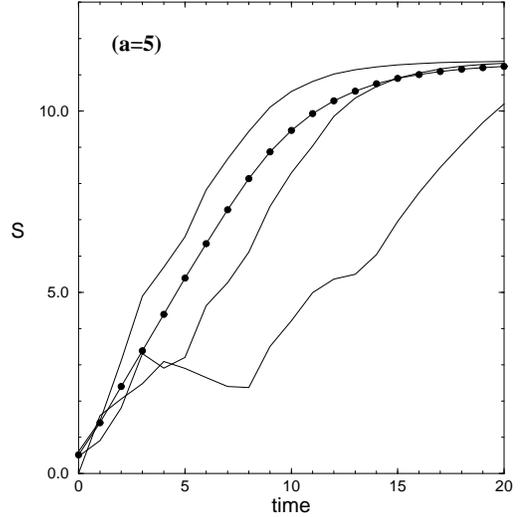,width=7truecm,angle=270}
\end{center}
\caption{ The evolution of $S(t)$ for the
Standard Map, with ~$k=5$. Also in this case the linear slope 
of the average curve (full circles) 
corresponds to the Kolmogorov--Sinai entropy ~$\kappa=0.98$.
Three single histories (thin full lines) are compared with the average one 
calculated over 1000 histories. See text for further details.
}
\end{figure}
%%%%%%%%%%%%%%%%%%%%%%%%%%%%%%%%%%%%%%%%%%%%%%%%%%%%%%%%%%%%%%%%%%%%%%%%%%%
We calculated numerically the Kolmogorov--Sinai entropy 
from the Lyapunov exponent, leaving out the regular islands 
for $a=5$. This yielded $\kappa=\lambda=$ 0.98,~1.62,~2.30, 
respectively for the three $a$'s. 
Each curve has a stage--2 linear portion whose slope is correctly 
given by $\kappa$. 
%%%%%%%%%%%%%%%%%%%%%%%%%%%%%%%%%%%%%%%%%%%%%%%%%%%%%%%%%%%%%%%%%%%%%%%%%%%
%%%%%%%%%  FIG. 5
\begin{figure}
\begin{center}
\epsfig{figure=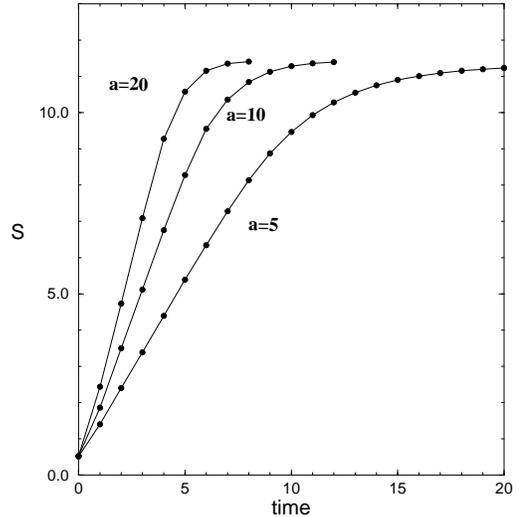,width=7truecm,angle=270}
\end{center}
\caption{ The same as fig.2 for the Standard Map,~$k=20,10,5$~(from right to left). In this case the Kolmogorov--Sinai entropy is 
~$\kappa=2.30,~1.62,~0.98$ respectively. It was considered 
~$N=10^6$,
~grid $=400\times400$,
~$V_i=V_{\rm cell}$. An
average of 100 trajectories  was used for $k=20,10$ , 
while 1000 histories were necessary  for $k=5$.
}
\end{figure}
%%%%%%%%%%%%%%%%%%%%%%%%%%%%%%%%%%%%%%%%%%%%%%%%%%%%%%%%%%%%%%%%%%%%%%%%%%%
We return again to the need for averaging 
many histories starting from different parts of phase space.  
This has to be done whenever the local Lyapunov exponents varies
appreciably from place to place, which is the normal case for nonlinear
systems. 
For linear maps (like the generalized cat map below) it is not
necessary. For other systems, it would never be necessary if we could use a
fine--enough coarse--graining, to give the probability time to spread
throughout phase space before any appreciable increase in entropy.
Unfortunately such fine grain would require computers far more
 powerful than exist now.  In the real thermodynamical
 world with many dimensions, what
kind of coarse--graining should preferably be used is, 
we believe, a wide open question.

%%%%%%%%%%%%%%%%%%%%%%%%%%%%%%%%

\section{The logistic map}

We focus now on the dissipative case, and more specifically 
on the logistic map:
\begin{equation}
X =1-a x^2\;\;\; -1 \le x \le1; \;0\le a \le2;\;\; 
\end{equation}
Despite being an extremely simple one--dimensional equation, 
the logistic map has always been used as a paradigmatic example of 
non--conservative chaotic systems, because it inglobes all the 
fundamental characteristics of a non--conservative system. 
The logistic map shows different regimes, according to 
the value of the control parameter $a$ \cite{beck}. 
In particular it is regular (negative Lyapunov exponent $\lambda$) for  
$a$ smaller than a critical value $a_c=1.401155198...$, and is 
chaotic for most part of the region $a>a_c$. 
At the chaos threshold $a=a_c=1.401155198...$ the Lyapunov 
exponent vanishes and this is the famous {\it edge of chaos} 
situation \cite{SOC}.
Before moving to the study of the entropy time evolution we need to 
discuss first the problem of the sensitivity to initial 
conditions.  
For all the cases in which the Lyapunov exponent $\lambda$ is positive  
we expect on the average an exponential increase of any small initial distance 
$ \xi(t) \equiv \frac{x_t-x_{t}^{\prime}}  {x_0-x_{0}^{\prime}}     $ 
to hold:
%%%%%%%% 
\begin{equation}
\label{lya}
\xi(t)=\exp{(\lambda t)}
\end{equation}
%%%%%
where $x_t$ and $x_{t}^{\prime}$ are the position 
at time $t$ of two initially close trajectories.
This case will be referred in the following as {\it strongly sensitive} 
to the initial conditions.
Instead at the 
edge of chaos $\lambda=0$, and the following equation has been 
proven to be valid in ref.~\cite{tsallisplastino} 
\begin{equation}
\xi(t)=[1+(1-q)\lambda_q t]^{\frac{1}{1-q}}\;\;\;(q \in \cal{R})
\end{equation}
which recovers Eq. (\ref{lya}) as the $q=1$ particular case. 
The case $q<1$ will be referred to as {\it weakly sensitive} to 
the initial conditions.  
So, {\it strong} and {\it weak} respectively stand for
{\it exponential} and {\it power-law} time evolutions of $\xi(t)$. 
A {\it weakly sensitive} system is well described once $q$, i.e.~ the 
exponent of the power--law, is given.
In particular at $a=a_c$ a value $q^*=0.2445...$ 
is obtained\cite{tsallisplastino}. 

We now apply to the logistic map the same kind of analysis used for 
conservative systems. The phase space interval $-1 \le x \le 1$ is 
partitioned into $10^5$ equal cells $c_{i}$. The initial 
distribution consists of $N=10^6$ points placed at random 
inside an interval picked at random anywhere on the phase space, and 
whose size is that of a cell. 
As the system evolves the probabilities $p_i(t)$ are computed 
at each time step.  

In order to study in the same framework either the chaotic case 
and the edge of chaos situation we consider a generalized 
non--extensive entropy proposed a decade ago in ref.~\cite{tsallis1} 
as a physical starting point to generalize statistical mechanics 
and thermodynamic:
\begin{equation}
\label{stsallis}
S_q(t)\equiv \frac{1-\sum_{i} [p_i(t)]^q}{q-1}\;\;\;\;(q \in \cal{R})\; .
\end{equation} 
This entropy is a function of the entropic index $q$ and 
reduces to the Boltzmann-Gibbs entropy, defined in equation 
(\ref{scoarse}), when $q=1$. 
A complete review of the existing theoretical, experimental
and computational work about this entropy can be found in ref. 
\cite{tsallis2}.
In particular it has been shown that such an entropy 
covers some types of anomalies due to a possible multifractal 
structure of the relevant phase space.  
For example, whenever we have long-range interactions 
\cite{anteneodo}, long-range microscopic memory \cite{buiatti}, or 
multifractal boundary conditions \cite{tsallisplastino}. 
%%%%%%%%%%%%%%%%%%%%%%%%%%%%%%%%%%%%%%%%%%%%%%%%%%%%%%%%%%%%%%%%%%%%%%%%%%%
%%%%%%%%%  FIG. 6
\begin{figure}
\begin{center}
\epsfig{figure=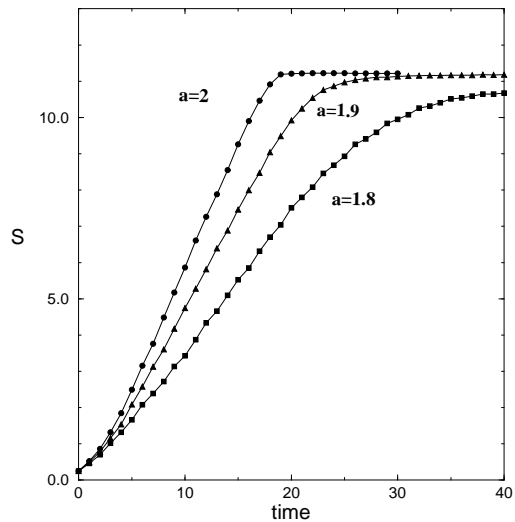,width=7truecm,angle=270}
\end{center}
\caption{ We show for the logistic map the 
time evolution of $S$ for the following values of $a$: 
$a=2$, $a=1.9$ and $a=1.8$~;
The curves are averages of 500 histories. The
slopes of the curves in the linear regime  are equal to the corresponding 
Lyapunov exponents. See text.}
\end{figure}
%%%%%%%%%%%%%%%%%%%%%%%%%%%%%%%%%%%%%%%%%%%%%%%%%%%%%%%%%%%%%%%%%%%%%%%%%%%
\noindent
We discuss in the rest of the paper the following results
obtained for the logistic map \cite{barangertsallis}:

\noindent
1) In the chaotic regime $q=1$: the Boltzmann-Gibbs 
entropy \ref{scoarse} exhibits a linear increase in time, and 
the rate of increase is equal to $\kappa$. 

\noindent
2) The standard Boltzmann-Gibbs entropy is inadequate to 
describe the edge of chaos situation. Instead it is 
the non--extensive entropy which grows linearly with time 
for a particular value $q\ne 1$.

\subsection{The chaotic case}

In fig. 6 we present our results for the logistic map in correspondence 
of three values for  $a$, all of them  in the 
chaotic regime, namely $a=2, 1.9, 1.8 $~.
Fluctuations are of course present (as $t$ increases), 
though their numerical importance can be cancelled by 
considering averages on the initial conditions.
Each of the curves is an average over 500 runs, i.e., 500 histories with
different initial distributions chosen at random, as mentioned. 
Though the asymptotic value, corresponding to a smooth 
distribution in the available part of phase space, is different in the 
three cases, all the curves show a linear increase on entropy. 
The slope in the intermediate time stage does not depend
on the dimension of the cells and on the 
distribution size \cite{barangertsallis} and is equal to the 
 predicted Lyapunov exponent, respectively $\ln 2$, $0.61$ and $0.48$~. 
Therefore the same results  found for conservative maps hold also for 
the logistic map, though the latter is a nonconservative one. 
%%%%%%%%%%%%%%%%%%%%%%%%%%%%%%%%%%%%%%%%
%%%%%%%%%%%%%%%%%%%%%%%%%%%%%%%%%%%
%%%%%%%%%  FIG. 7
\begin{figure}
\begin{center}
\epsfig{figure=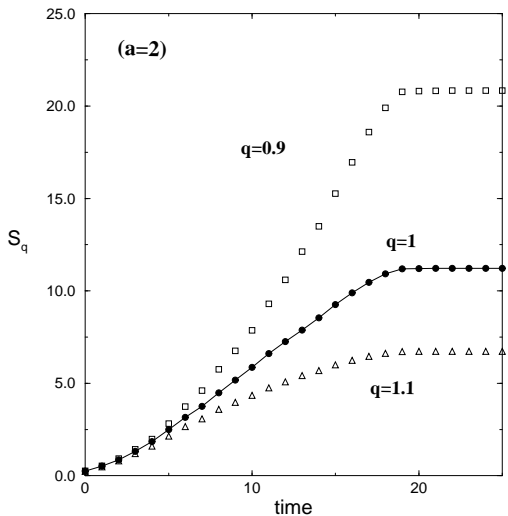,width=7truecm,angle=270}
\end{center}
\caption{ For the logistic map  with $a=2$,  we 
draw the time evolution of the non--extensive entropy $S_q$, considering 
three different values of $q$.  Results are averages over 500 histories.
See text for further details.}
\end{figure}
%%%%%%%%%%%%%%%%%%%%%%%%%%%%%%%%%%%%%%%%%%%%%%%%%%%%%%%%%%%%%%%%%%%%%%%%%%%
In fig.~7 we focus on the case $a=2$ and instead of  
$S(t)$ we consider the time 
evolution of $S_q$, as defined by eq. \ref{stsallis}, for three 
different values of $q$~. 
As $t$ evolves, $S_q(t)$ tends to increase (in all cases bounded by 
$\frac{M^{1-q}-1}{1-q}$, or $\ln M$ when $q=1$).  
Only the curve for $q=1$ shows a clear linear behavior and the  
slope is equal to the Lyapunov exponent $\ln 2$~.   
For $q<1$ the curve is concave, while for $q>1$ the curve is convex. 
Therefore for the logistic map in the chaotic regime the standard 
Boltzmann--Gibbs must be used as for any hamiltonian chaotic system
or conservative map.

\subsection{The edge of chaos}

So far we have shown that $q$ is 1 for all the cases in which the logistic
curve is chaotic, i.e. strongly sensitive to the initial conditions.  Now we
want to study the same system at its chaos threshold, i.e.  at  $a=a_c \simeq
1.401155198...$ for which the Feigenbaum attractor exists.  We expect in this case 
a particular value 
$q\ne 1$  , due to the fractality of phase space \cite{barangertsallis}  and power-law
sensitivity to initial conditions. 
For such a value of $a$, much bigger fluctuations than in the chaotic 
case are observed. To understand this it is sufficient to 
consider that the attractor occupies only 
a tiny part of phase space
(844 cells out of the $10^5$ of our partition).  
We therefore require a very efficient and careful 
averaging over the initial conditions. 
Here we adopt a selection method of the  best histories  
based on how good is each initial condition at 
spreading itself. We obtain 1251 histories and we 
address to ref. \cite{barangertsallis} 
for all the details about the selection method. 
In fig.8 we plot $S_q(t)$ for four different values of $q$~; 
the curves are an average over the 1251 histories selected. 
The growth of $S_q(t)$ is found to be
linear when $q=q_c=0.2445$, while for $q<q_c$ ($q>q_c$) the curve is concave
(convex).  
%%%%%%%%%%%%%%%%%%%%%%%%%%%%%%%%%%%%%%%%%%%%%%%%%%%%%%%%%%%%%%%%%%%%%%%%%%%
%%%%%%%%%  FIG. 8
\begin{figure}
\begin{center}
\epsfig{figure=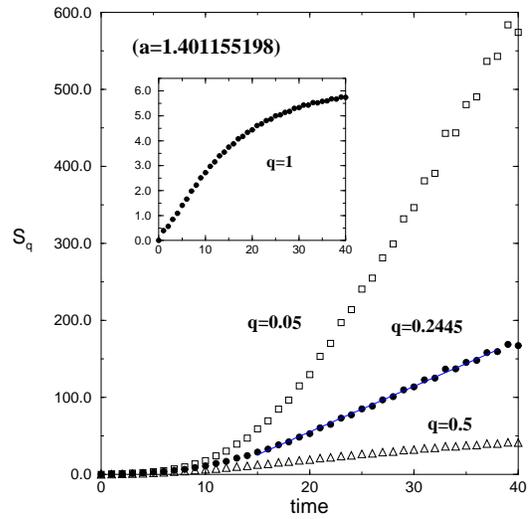,width=7truecm,angle=270}
\end{center}
\caption{ Time evolution of $S_q$ for $a=a_c$~.  
We consider four different values of $q$~. 
The case $q=1$ is in the inset.  Results are averages over 1251 histories.}
\end{figure}
%%%%%%%%%%%%%%%%%%%%%%%%%%%%%%%%%%%%%%%%%%%%%%%%%%%%%%%%%%%%%%%%%%%%%%%%%%%
This behavior is similar to the one in fig.7, with a major
difference: {\it the linear growth is not at $q=1$, (see inset in fig.8), 
but at $q=q_c=0.2445$.  } 
To extract the particular value of q for which we get the best linear rise of
the nonextensive entropy,
we have fitted the curves $S_q(t)$ in the time
interval $[t_1,t_2]$ with the polynomial $S(t)=a+bt+ct^2$~. We define 
the coefficient $R=|c|/b$ as a measure of the importance of the nonlinear
term in the fit: if the points are on a perfect straight line, $R$ should be
zero.  We choose $t_1=15$ and $t_2=38$ for all $q$'s. 
Fig.9 shows that the minimum of  $R=|c|/b$  occurs  for $q=q_c=0.2445$.

The value of q which allows for a linear growth 
of $S_q(t)$,
obtained through this procedure,  happens to coincide with the value 
$q^*$ obtained in a completely different method by studying the power-law 
sensitivity to initial conditions \cite{tsallisplastino}.

%%%%%%%%%%%%%%%%%%%%%%%%%%%%%%%%%%%%%
%%%%%%%%%%%%%%%%%%%%%%%%%%%%%%%%%%%%%%
%%%%%%%%%  FIG. 9
\begin{figure}
\begin{center}
\epsfig{figure=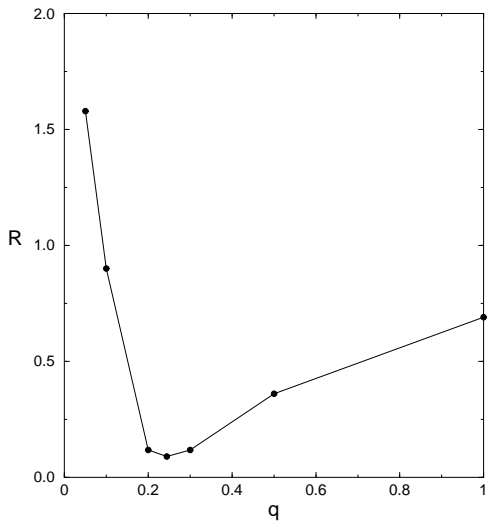,width=7truecm,angle=270}
\end{center}
\caption{ Coefficient of nonlinearity $R$ vs. $q$.~ See text for further details.}
\end{figure}
%%%%%%%%%%%%%%%%%%%%%%%%%%%%%%%%%%%%%%%%
%%%%%%%%%%%%%%%%%%%%%%%%%%%%%%%%%%%

There exists also another method, based only on the geometrical 
description of the multifractal attractor existing at $a_c$, 
which gives exactly the same value of $q$\cite{tsallisplastino}. 
The multifractal attractor can be characterized by 
using the multifractal function $f(\alpha)$
\cite{beck}. This function is defined in the interval
$[\alpha_{min},\alpha_{max}]$, and its maximum equals the fractal
or Hausdorff dimension $d_f$. 
For a large class of systems the value of $q$ can be calculated from
\begin{equation}
\frac{1}{1-q^*}= \frac{1}{\alpha_{min}}-\frac{1}{\alpha_{max}}~,
\end{equation}
where $\alpha_{min}$ ($\alpha_{max}$) is the lowest (highest) value for which
$f(\alpha)$ is defined.  
In particular for the logistic map at the edge of chaos 
$\alpha_{min}=0.380...$, $\alpha_{max}=0.755...$ and 
$q^*=0.2445$. 

Finally, it is also interesting to note, that our   
result for the linear growth of $S_q$ at the edge of chaos has been recently 
confirmed by using a different method
\cite{grigo}.

\section{Conclusions}

To summarize, we have illustrated, through several numerical examples
 for conservative and dissipative chaotic maps, the relationship existing between
the Boltzmann equilibrium thermodynamic entropy $S$ 
the Kolmogorov--Sinai one $\kappa$
used for dynamical systems. As Krylov already suggested
in the early '40s \cite{krylov}, the mixing property characteristic of 
chaos  is  fundamental 
in order to reach the thermodynamical equilibrium. When using 
a coarse-graining procedure, the growth of  $S(t)$ averaged over 
many histories is linear and the slope gives just the Kolmogorov--Sinai 
entropy.  This has been verified also for a dissipative case, i.e. 
the logistic map in the chaotic regime.
Finally,
 a  very  interesting generalization must  be done at the chaos threshold,
where the sensitivity to initial conditions is not exponential, but  power--law
like. In this case, in fact,  in order to have a linear growth for the entropy, 
as  for the full chaotic 
regime,  the non--extensive entropic form introduced by Tsallis should
be adopted.   The latter has been successfully  checked for many cases where
long--range correlations and a fractal phase space is found.  We have shown
that 
we get a linear  growth for the generalized entropy 
only when the value $q=0.2445$ (and not $q=1$ for which the standard
entropy is recovered) is used.  This  fact confirms
previous numerical studies and generalizes the connection between the two
entropies also at the edge of chaos.
We conclude hoping  that our results could stimulate a deeper 
undestanding of the connections between dynamics and statistical mechanics.
 
It is a pleasure to aknowledge the 
stimulating collaboration with Constantino Tsallis for the results 
on the logistic map.

\end{document}